\documentclass[a4paper,fleqn,nopreprintline]{cas-sc}
\usepackage[numbers]{natbib}
\usepackage{bm}% bold math
\usepackage[normalem]{ulem}
\usepackage{color}
\usepackage{graphicx}
\date{}
\usepackage{placeins}

\RequirePackage{marginnote}

\begin{document}
%\let\WriteBookmarks\relax
%\def\floatpagepagefraction{1}
%\def\textpagefraction{.001}

\shorttitle{}

%\shortauthors{<short author list for running head>}

\title [mode = title]{Rashba-induced spin texture and spin-layer-locking effects  in the antiferromagnetic CrI$_{3}$ bilayer}

%\tnotemark[<tnote number>]

%\tnotetext[<tnote number>]{<tnote text>}

% eg: \author[1,3]{Sukanya Ghosh}[type=editor,

\author[]{Sukanya Ghosh\footnote{Present affiliation: Department of Physics and Astronomy, Uppsala University, PO Box 516, SE-751 20 Uppsala, Sweden\thispagestyle{empty}
\\
*email: sukanya.ghosh@physics.uu.se}}

\cormark[1]

%\fnmark[1]{sukanya.ghosh@physics.uu.se}

%\ead{sukanya.ghosh@physics.uu.se}

%\ead[url]{<URL>}

%\credit{<Credit authorship details>}


%\author[<aff no>]{<author name>}[<options>]

% Footnote of the second author

% Email id of the second author
%\ead{}



%\affiliation[<aff no>]{organization={},
%          citysep={}, % Uncomment if no comma needed between city and postcode

%\cortext[1]{Corresponding author}


%\author{A. Author}
% \altaffiliation[Also at ]{Physics Department, XYZ University.}%Lines break automatically or
%\author{Sukanya Ghosh\footnote{Present affiliation: Department of Physics and Astronomy, Uppsala University, Box-516, 75120 Uppsala, Sweden}}

 %\altaffiliation[Also at ]{International Centre for Theoretical Physics, Trieste, Italy}%Lines break automatically or can be forced with \\
\author{Nata{\v s}a Stoji{\'c}}%
\author{Nadia Binggeli}
\address[1-3]{Abdus Salam International Centre for Theoretical Physics,  Strada Costiera 11, 34151 Trieste, Italy}
%\affiliation{%
% Abdus Salam International Centre for Theoretical Physics,  Strada Costiera 11, Trieste, Italy
%\footnotetext{uuuu}

%\email{sukanya.ghosh@physics.uu.se}



\begin{abstract}
  The antiferromagnetic (AFM) CrI$_3$ bilayer is a particularly interesting representative of van der Waals 2D semiconductors, which are currently being studied for their magnetism and for their potential in spintronics. Using ab initio density-functional theory calculations,
we investigate the spin texture in momentum space of the states of the (doubly degenerate) highest valence band of the AFM CrI$_3$ bilayer with Cr-spin moments perpendicular to the layers. We find  the existence, in the main central part of the Brillouin zone, of a Rashba in-plane spin texture of opposite signs on the two layers,  resulting from the intrinsic local electric fields acting on each layer.
To study the layer segregation of the wavefunctions,
we apply a small electric field that splits the degenerate states according to their layer occupancy. We find  that the wavefunctions of the highest valence band are layer-segregated, belonging to only one of the two layers with opposite in-plane spin textures, and the  segregation occurs  over nearly the whole Brillouin zone.
The corresponding layer locking of the in-plane-canted spin is related to  the separation in energy of the highest AFM band from the rest of the valence bands. We explain how the band interactions destroy the layer locking  at the $K$, $K'$, and $\Gamma$ points. Furthermore, we compare the layer locking of the in-plane-canted spin in our AFM bilayer system with the hidden spin polarization in centrosymmetric nonmagnetic materials, pointing out the differences in segregation mechanisms and their consequences for the layer locking.
 We note that a similar Rashba  effect with layer locking of in-plane-canted spin could occur in other van der Waals AFM bilayers with strong spin-orbit coupling and an isolated energy band.

\end{abstract}


%\item
%\item
%\item

% Keywords

\maketitle

% Main text
%\section{}\label{}

% Use the style of numbering in square brackets.
% If nothing is used, default style will be taken.
%\begin{enumerate}[a)]
%\item
%\item
%\item
%\end{enumerate}

%\begin{itemize}
%\item
%\item
%\item
%\end{itemize}

%\begin{description}
%\item[]
%\item[]
%\item[]
%\end{description}

% Figure
%\begin{figure}[<options>]
%	\centering
%		\includegraphics[<options>]{}
%	  \caption{}\label{fig1}
%\end{figure}

%Use showkeys class option if keyword
                              %\maketitle
%\tableofcontents

%\keywords{aaa}%Use showkeys class option if keyword

\section{Introduction}
%\protect\\ \lowercase{via} \textbackslash\textbackslash

\pagestyle{empty}

The recent discovery of intrinsic magnetic order in atomically thin semiconducting van der Waals crystals \cite{GonLiLi17,HuaClaNav17} is  opening avenues for   fundamental research into 2D magnetism \cite{BurManPar18} and for the creation of  low-power spintronic devices \cite{LinYanWan19}. A particularly interesting  material in this context is CrI$_3$, in its monolayer and bilayer forms. Monolayer CrI$_3$ is ferromagnetic (FM)  with an out-of-plane magnetization easy axis and a Curie temperature of 45~K \cite{HuaClaKle18}. In a bilayer, the weak van der Waals forces bind the two monolayers, whose intralayer couplings remain FM. The CrI$_3$ bilayer is  experimentally observed to be a layered antiferromagnet \cite{HuaClaNav17,UbrWanTey19,ThiWanTsc19,JiaShaMak18,HuaClaKle18}, with a N{\'e}el temperature similar to the monolayer's Curie temperature \cite{JiaShaMak18,HuaClaKle18}.

In the last decade, antiferromagnets have become increasingly more appealing for spintronics, owing to their advantageous properties with respect to ferromagnets:  antiferromagnetic (AFM) materials do not produce stray fields, permitting high-density memory integration, and are much less sensitive to
magnetic field perturbations, providing the necessary stability for data storage \cite{YunMarWad16,BalManTso18}. In addition, antiferromagnets have much faster spin
dynamics than ferromagnets, which is necessary  for ultrafast data processing \cite{LebRosBen18}.

The applications of the CrI$_3$ bilayer are conceived on the basis of its layered AFM ground state and the ease of its transformation to the FM phase (low  critical magnetic field of 0.6--0.7~T) \cite{HuaClaNav17}.
The switching  and tuning of the interlayer exchange from AFM to FM has been heavily investigated and  obtained experimentally by various external perturbations \cite{JiaShaMak18,HuaClaKle18,JiaXieSha20,Song_2019}.   For example,  the CrI$_3$ bilayer and its multilayers are  unique in their ability to  simultaneously act as a  tunneling barrier and a spin-filtering layer, thanks to their semiconducting nature and switchable magnetic state. Consequently, they display exceptional  magnetoresistance properties  upon the application of a magnetic field  in both
experimental studies \cite{KleMacLad18,SonCaiTu18,KimYanPat18} and theoretical studies \cite{Tsymbal_2019,YanCaoJia21}. Recently, proximity spin-orbit torque on the CrI$_3$ bilayer in a van der Waals heterostructure was shown theoretically \cite{DolPetZol20}. At the same time, giant second-harmonic generation was observed in the AFM CrI$_3$ bilayer and was explained by the simultaneous absence of both inversion symmetry and time-reversal symmetry in its layered antiferromagnet structure \cite{SunYiSon19},  which are also found to activate and control Raman divergent optical selection rules \cite{HuaCenZha20}. The influence of symmetry operations for the magneto-optical Kerr effect was recently shown for AFM and FM CrI$_3$-CrBr$_3$ and CrI$_3$ bilayers \cite{YanHuWu20}. However,
despite the great interest in the AFM ground state of the bilayer and newly discovered properties and phenomena stemming from its layered magnetic order, little is known about its {\bf k}-space spin texture. Spin texture can be of vital importance as it determines how spin-polarized currents can be manipulated for spintronic devices. It has been heavily investigated in nonmagnetic materials, especially after the discovery of hidden spin polarization \cite{ZhaLiuLuo14,YuaLiuZha19,TuCheRua20,CheSunChe18} and spin-layer locking \cite{YaoWanHua17,ZhaZhaHao21} in inversion-symmetric nonmagnetic semiconductors and semimetals.
In contrast,  the influence of the AFM order and the underlying symmetries on the spin texture in general is not well understood and is currently being studied in other systems, such as in the spin textures in hybrid organic–inorganic perovskites \cite{LouGuJi20}.

In this work, we study the {\bf k}-space spin texture of the CrI$_3$ bilayer upper valence states by using density-functional theory. We find that the spin texture of the bilayer is  essentially tangential and characterized by layer locking of the in-plane-canted spin. In addition,
the Rashba effect is not accompanied by the typical energy splitting. To be able to evaluate the layer-dependent properties, we apply a small electric field that splits the AFM degenerate states according to their layer localization. We find that the wavefunctions of the two highest  valence states are spatially segregated almost   throughout the Brillouin zone (BZ) on one of the two layers featuring opposite in-plane spin textures. We point out that the wavefunction layer segregation is made possible  by the AFM order and energy separation of the band. We show that the layer segregation is broken essentially only at the $K$ and $K'$ points and somewhat at the $\Gamma$ point, which can be understood on the basis of the properties of monolayer band structure. We also compare the Rashba effect in the AFM bilayer with the hidden spin polarization in centrosymmetric nonmagnetic 2D materials.

\section{System and computational method}
%\protect\\ \lowercase{via} \textbackslash\textbackslash

Our calculations are performed using density-functional theory as implemented in the Quantum ESPRESSO package with the plane-wave basis set and pseudopotentials \cite{GiaBarBon09}.
The local spin density approximation is used as the exchange-correlation functional to treat interactions between the electrons \cite{Perdew_Zunger}.
The coupling between electron spin and its orbital angular momentum is considered by our including the fully relativistic effect. The electron-ion interactions are described by the fully relativistic projector augmented wave  pseudopotentials \cite{PhysRevB.50.17953}.
The plane-wave cutoffs for the kinetic energy and charge density for all our calculations are 60~Ry and 650~Ry, respectively. We use the $\Gamma$-centered  Monkhorst-Pack $k$-point grid of  $24 \times 24 \times 1$.
The periodic images of the CrI$_3$ bilayer are separated by a vacuum region with  a thickness of 30~\AA.
In the structural relaxations, all coordinates are relaxed until the force on each atom becomes smaller than 0.1~mRy/bohr.

We concentrate on the CrI$_{3}$ bilayer with monoclinic layer stacking and $C_{2h}$ space-group symmetry. This is the common experimental form of the bilayer obtained by exfoliation from bulk CrI$_3$ at room temperature, and which remains in that form when cooled to temperatures below the N\'eel temperature \cite{SunYiSon19,UbrWanTey19}. The bilayer in that structure has been the focus of enormous attention because of its layered AFM ground state, which can be conveniently switched experimentally to the FM state (by a weak magnetic field as well as by electrical means) \cite{HuaClaNav17,HuaClaKle18,JiaShaMak18,JiaXieSha20}.

 Our optimized equilibrium local density approximation lattice constant for the CrI$_3$ bilayer is  6.69~\AA, and the innermost distance between the two layers is 3.36~\AA\ \cite{GhoStoBin21}, in good agreement with the findings of a previous local density approximation study \cite{Leon_2020}.
 The corresponding AFM configuration is energetically more stable than the FM configuration \cite{GhoStoBin21}, consistent with the experimental situation  \cite{HuaClaNav17,UbrWanTey19,ThiWanTsc19,JiaShaMak18,HuaClaKle18}. Our calculated value of the Cr atomic spin moment of 2.79~$\mu_\mathrm{\mathrm{B}}$ for bilayer and monolayer CrI$_3$ also closely corresponds to the experimental values \cite{FriDufZha18,McGDixCoo15}, and the monolayer magnetic anisotropy energy $E_{\parallel} - E_{\perp} $ of 0.77~meV/Cr places the magnetization in the perpendicular  direction, as found  experimentally \cite{FriDufZha18,CheChuChe20}, and in good agreement with the findings of previous theoretical studies (see Ref.~\cite{Leon_2020} and references therein).
 %{\Ndel{We note that our calculated Cr magnetic moment of 3~$\mu_B$/Cr for monolayer and bilayer CrI$_3$ corresponds closely to the experimental values, \cite{McGDixCoo15,FriDufZha18} while the monolayer magnetic anisotropy energy $E_{\parallel} - E_{\perp} $} of 0.65~meV/Cr places the magnetization in the perpendicular direction,\cite{FriDufZha18,CheChuChe20} as found  experimentally, and is in  good agreement with a previous LDA study.\cite{Leon_2020}}   }

 The band structure and the spin texture of the AFM bilayer are studied  in the presence of an external electric field applied  perpendicularly to the bilayer plane (positive $z$ direction). The electric field is modeled with a sawlike potential along the $z$ direction of the bilayer supercell. Dipole correction is applied to avoid spurious
interactions between the periodic images of CrI$_{3}$ along the direction of the external electric field \cite{Lennart_1999,SG_2019}. The layer-projected band structure and the corresponding weights of the states on each layer are obtained from the atomic pseudo-orbital  projections of the states for the atoms in the layer.

\begin{figure}[t]
\centering
\includegraphics[width=0.7\textwidth] {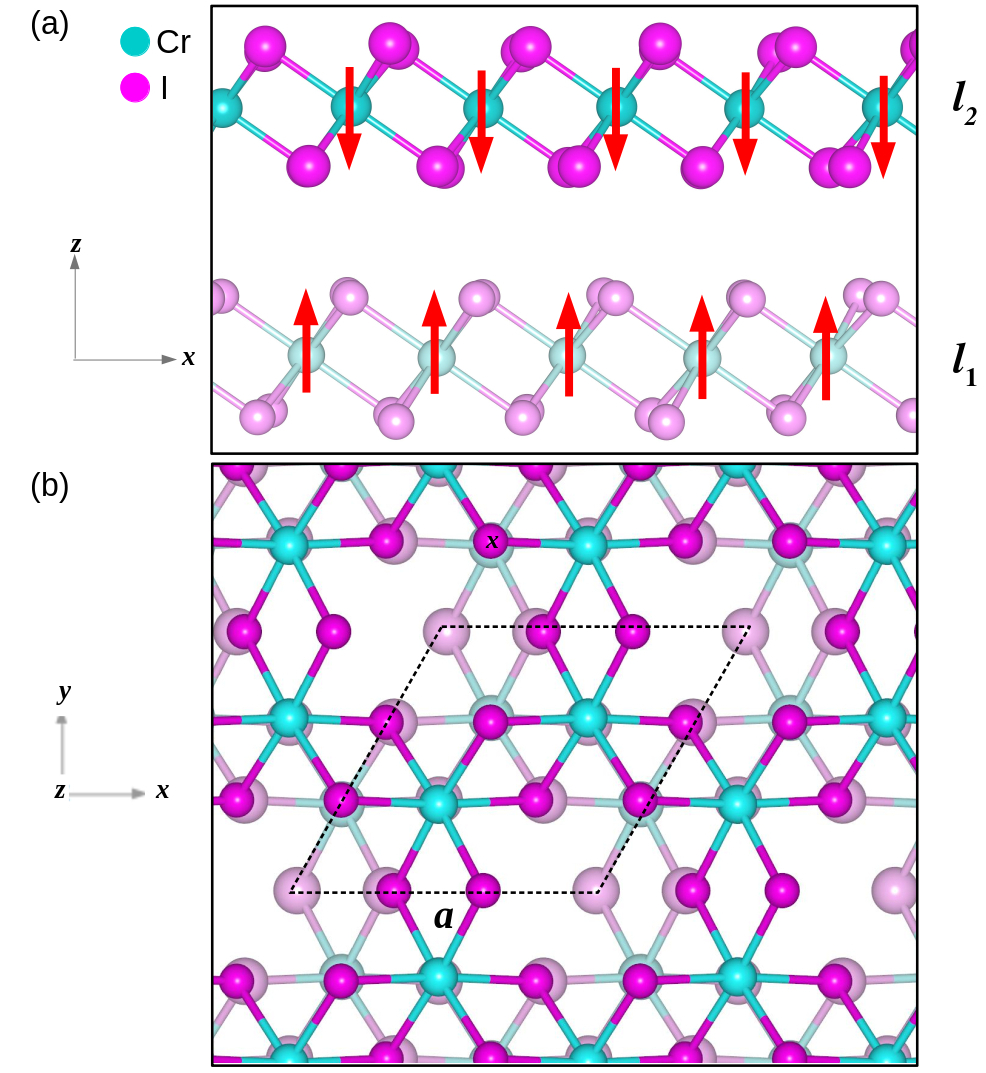}
\caption{\label{structure_AFM} (a) Side view of the atomic structure of the antiferromagnetic CrI$_{3}$ bilayer with monoclinic layer stacking. In each CrI$_{3}$ monolayer, the plane of Cr atoms is sandwiched between two I atomic planes. The turquoise and magenta spheres represent the Cr and I atoms, respectively. The magnetic moments on the Cr atoms are denoted by the red arrows, while $l_1$ and $l_2$ stand for the two layers. (b) Top view of the CrI$_{3}$ bilayer, where the dashed black  lines show the unit cell and \textit{a} is the lattice parameter.}
%The external electric field $E_\mathrm{field}^\mathrm{ext}$ in the AFM bilayer is applied along the out-of-plane direction shown by the vertical black arrow.
\end{figure}

 %{\NBdel {In this study we consider monoclinic layer stacking in CrI$_{3}$ bilayer, with C$_{2h}$ point-group symmetry of the atomic structure (without electric field) and AFM interlayer coupling. }}
%In this study we consider monoclinic layer-stacking in CrI$_{3}$ bilayer which belongs to the magnetic point group C$_{2h}$ and the space group C2/m.
 %This is the experimentally observed structural phase at low temperature, and achieved by exfoliation from CrI$_3$ bulk van der Waals crystals at room temperature.\cite{HuangNature2017,Ubrig_2019,Thiel973,Jiang_2018,Huang_2018}
 Fig.~\ref{structure_AFM} shows the side and top views of the CrI$_{3}$ bilayer in the AFM configuration; the arrows denote the magnetic ordering on the two layers, $l_1$ and $l_2$.  The corresponding magnetic point group is $C_{2h}$ ($C_2$), with  the following four symmetry operations: (i)  identity E; (ii) 180$^\circ$ rotation about the $y$ axis R$_{y}$[180$^\circ$]; (iii) inversion combined with time reversal I$\cdot$T; and $x$-$z$ mirror-plane reflection combined with time reversal  M$_{xz}\cdot$T. The presence of the  I$\cdot$T symmetry implies that all bands of the pristine AFM bilayer are doubly degenerate (or multiples of that at some {\bf k} points); the pristine band structure is presented in Ref. \cite{GhoStoBin21}). The presence of the electric field breaks the I$\cdot$T symmetry and lifts that degeneracy (and also breaks the  R$_{y}$[180$^\circ$] symmetry).

 %CrI$_{3}$ bilayer in AFM configuration with magnetic point group C$_{2h}$ possesses the following symmetry operations: i) identity and ii) inversion ($\mathcal{P}$) followed by $180^{0}$ rotation about $y$ together with time-reversal symmetry (T).

We obtain the spin texture  of the band states in {\bf k}-space by calculating the expectation values of the spin operators $\frac{1}{2}\hat{\sigma_{\alpha}}$, $\alpha=x,y,z$, where  $\hat{\sigma_{\alpha}}$ are the Pauli matrices.
The spin components corresponding to the $i$th energy spinor eigenfunction, $\Psi_{i, {\bf k}}({\bf r})$, are obtained as

$S_{\alpha} ({\bf k})=  \frac{1}{2} \frac{<\Psi_{i,{\bf k}}|\hat{\sigma}_{\alpha}|\Psi_{i, {\bf k}}>}{<\Psi_{i, {\bf k}}|\Psi_{i, {\bf k}}>}$.

The particular calculation procedure  to obtain the spin textures is given in the supplementary material.

We also examine the magnetization density, {\bf m}({\bf r}), of some of the band states at selected {\bf k} points. The magnetization density components are evaluated as $m_{\alpha} ({\bf r}) = \mu_\mathrm{B} \Psi^{+}_{i,{\bf k}}({\bf r}) \hat{\sigma}_{\alpha} \Psi_{i,{\bf k}}({\bf r})$, where $\mu_\mathrm{B}$ is the Bohr magneton. When the states in the band are degenerate, as is the case in the AFM bilayer without an electric field, we evaluate the magnetization density components by summing over the degenerate states:

$m_{\alpha} ({\bf r}) = \mu_\mathrm{B} \displaystyle\sum_{i \in {\rm band }   \,n}  \Psi^{+}_{i,{\bf k}}({\bf r}) \hat{\sigma}_{\alpha} \Psi_{i,{\bf k}}({\bf r}) $.

\section{Results and discussion}

\begin{figure*}[t]
\centering
\includegraphics[width=0.7\textwidth] {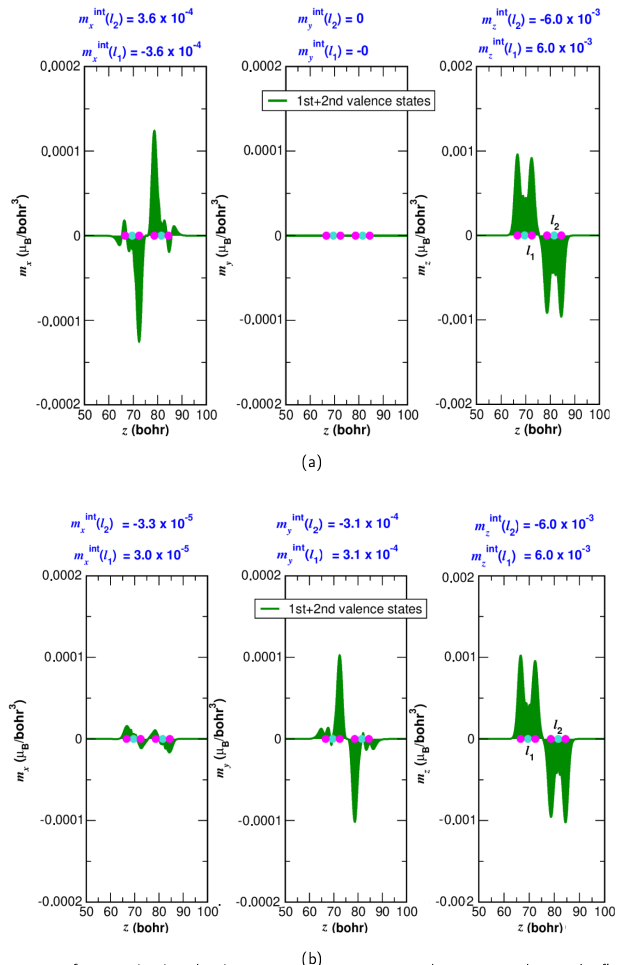}
\caption{\label{fig:magn_ky_kx} Planar average  of magnetization-density components   $m_{x}$, $m_{y}$, and $m_{z}$ summed over the first and second (the two highest) valence states of the pristine CrI$_{3}$ bilayer in the antiferromagnetic configuration at (a) ${\bf k} = (0, 0.15) \times 2 \pi/a$ and (b) ${\bf k} = (0.13, 0) \times 2 \pi/a$. The turquoise and magenta circles show the positions of Cr and I atoms, respectively, along the $z$ direction of the supercell, and $l_{1}$  and $l_{2}$ denote  the CrI$_3$ layers in the bilayer. These two states contribute to form the highest valence band of the antiferromagnetic bilayer. $m_{i}^{\rm int}(l_{1})$ and $m_{i}^{\rm int}(l_{2})$, $i=x,y,z$,  are obtained by integrating $m_{i}$ on $l_{1}$ and $l_{2}$, respectively, and are given in the unit of $\mu_{\mathrm{B}}/$bohr$^{2}$. } % caption for whole figure
%\end{center}
\end{figure*}

In Fig.~\ref{fig:magn_ky_kx}, we show the planar average of the  magnetization density components $m_x(z), m_y(z)$, and $m_z(z)$ for the highest valence band of the pristine AFM bilayer at two {\bf k} points, chosen at similar distances from the BZ center along the $k_y$ axis [Fig.~\ref{fig:magn_ky_kx}(a)] and along the $k_x$ axis [Fig.~\ref{fig:magn_ky_kx}(b)]. The magnetization density was obtained as the sum of the magnetization densities of the two degenerate states of the band at those {\bf k} points.

Apart from the $m_z$ component parallel to the Cr-layer spin, which is virtually the same for the two {\bf k} points, we observe the presence of an in-plane magnetization-density component that strongly depends on the direction and length of the {\bf k} wavevector. The amplitude of this in-plane magnetization-density component is about one tenth that of the $m_z$ component.
For the {\bf k} point along the $k_y$ axis, ${\bf k} = (0, 0.15)\times 2\pi/a$, in Fig.~\ref{fig:magn_ky_kx}(a), $m_x$ is significant and displays identical magnetization of opposite sign on the two layers of the bilayer, $l_1$ and $l_2$, while  $m_y$ remains zero on both layers. The  $m_z$ component also changes sign on the two layers, following the sign of the Cr spins on the two layers.
For the {\bf k} point along the $k_x$ axis, in Fig.~\ref{fig:magn_ky_kx}(b), $m_y$ is also significant and inverts its sign from $l_1$ to $l_2$, while keeping the same amplitude. In this case, the  $m_x$ component is not vanishing, but is of much smaller magnitude than the $m_y$ component.

 Thus, in both cases
 the magnetization-density component parallel to the ${\bf k}$ vector is either zero ($m_y$ in Fig.~\ref{fig:magn_ky_kx}(a)) or much smaller than the  in-plane tangential component ($m_x$ in Fig.~\ref{fig:magn_ky_kx}(b) is one order of magnitude smaller than $m_y$).
The behavior of the in-plane tangential magnetization-density components for both wavevectors is fully consistent with the tangential spin texture produced by the Rashba effect for electric field  ${\bf E}$ perpendicular to the layer, and corresponding effective magnetic field $\mathbf{B}_{\rm eff} \sim {\bf k} \times {\bf E}$. In the AFM CrI$_3$ bilayer, each layer feels an electric field originating from the presence of the other layer.  The fields acting on the two layers are oppositely oriented (we estimate the strength of this intrinsic electric field to be on the order of 0.1~V/\AA\  at  each layer; see the supplementary material). This creates tangential  Rashba in-plane spin textures of opposite sign on the two layers.

For ${\bf k}$ along the $k_x$ axis, there is also a very small radial $m_x$ magnetization-density component present on each layer. This originates in the local chirality of the FM monolayer (i.e., its lack of horizontal mirror-reflection symmetry), an effect discussed in detail in Ref.~\cite{FM_SG} for the FM bilayer. In the latter case, apart from this local chirality, the FM bilayer is also globally chiral (unlike the AFM bilayer), which results in dominant radial components along the $k_x$ axis and equivalent directions of the FM bilayer \cite{FM_SG}. For the AFM bilayer, the simultaneous tangential behavior and opposite signs of the magnetization-density components for the two layers at the two {\bf k} points (and also observed at other {\bf k} points) suggest a layer locking and spatial segregation of  the degenerate states of the upper valence band  with opposite helicities of the in-plane spin texture in the two  layers.

\begin{figure*} [pos=t]
\centering
\includegraphics[width=0.75\textwidth] {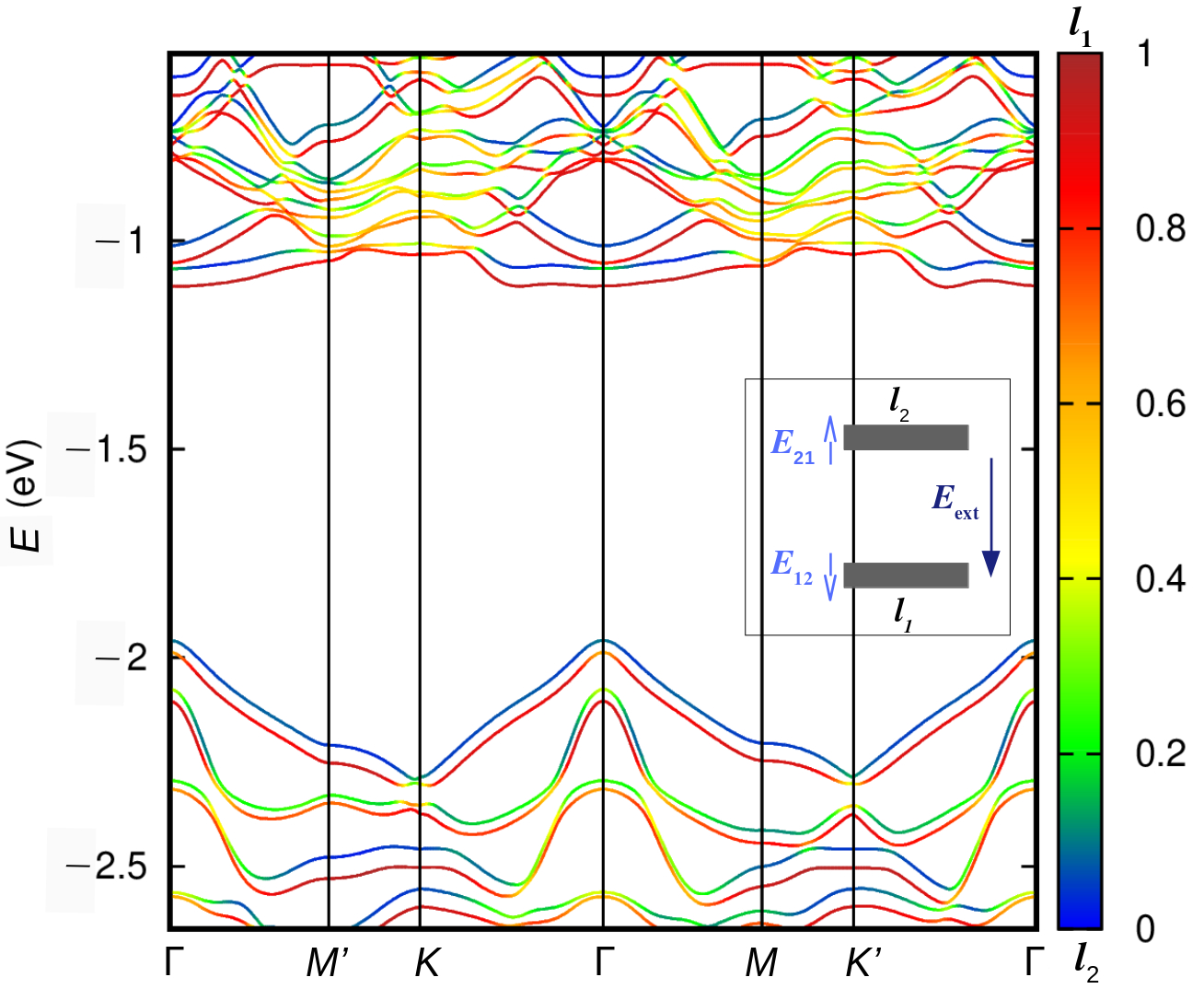}
\caption{\label{fig:band_Efield} Band structure of the antiferromagnetic CrI$_{3}$ bilayer in the presence of an external electric field, $E_{\rm ext} = 0.05$~V/\AA, applied along the out-of-plane direction of the CrI$_{3}$ bilayer.  Also shown is the color-coded weight of the states on layer 1 ($l_\mathrm{1}$) of the  antiferromagnetic bilayer. The energy window includes the highest
valence bands and the lowest conduction bands. The inset presents a schematic side view of the bilayer unit cell with the two monolayers $l_\mathrm{1}$ and $l_\mathrm{2}$, where the light blue arrows denote the local internal electric fields $E_\mathrm{12}$ and $E_\mathrm{21}$ present in the bilayer and the dark blue arrow denotes the external electric field $E_\mathrm{ext}$.}
\end{figure*}

To confirm the layer segregation of the states with opposite in-plane-canted spins, we applied a  small electric field, $E_{\mathrm{ext}} =  0.05$~V/\AA, along the $z$ direction of the unit cell, and evaluated the corresponding layer-projected band structure and corresponding layer-weighted spin texture.
%which splits the doubly degenerate energy levels of the AFM bilayer.
The electric field places   each of the bilayer's layers at a different potential and thus splits the AFM degenerate energy levels according to the state's localization on $l_1$ and $l_2$.
The effect of this electric field on the CrI$_3$ bilayer band structure is shown in Fig.~\ref{fig:band_Efield}.
 There, we also display the weight of every state on  $l_1$. Importantly, the highest valence  band is uniformly split nearly   throughout the BZ, with a weight of the lower (upper) split state of more than 90\% (less than 8\%) on $l_1$. In other words, the two states belonging to the highest band are layer-separated: one is segregated  on $l_1$ and the other is segregated on $l_2$. It is important to underline that the electric field separates the states only  in energy, while the spatial layer segregation of the states is a property of the pristine system. The only exceptions from the spatial segregation of the two highest states are around the special {\bf k} points $K$ and $K'$, at which each state has a noticeable weight on both $l_1$ and $l_2$. The highest valence states at the $K$ and $K'$ points are rather special already in the monolayer (where the states are degenerate; see the Appendix), and upon creation of the AFM bilayer, they interact, hybridizing and creating an energy splitting. To a smaller extent, also around $\Gamma$, the two highest valence states (especially the second-highest valence state), in Fig.~\ref{fig:band_Efield}, are not completely layer-locked.  This can be similarly understood on the basis of the monolayer band structure (Fig.~\ref{fig:ML_Sz} in the Appendix); there, at the $\Gamma$ point, the states of the two highest valence bands are particularly close in energy  and tend to exhibit increased hybridization upon formation of the AFM bilayer.   In  Fig.~\ref{fig:band_Efield},  we can see that the other bands, both conduction and valence bands, are not uniformly split in general and that many are characterized by similar weights on the two layers.  The upper valence band of the pristine AFM CrI$_3$ bilayer \cite{GhoStoBin21} has the particular property of being separated in energy from the other bands. This separation somewhat decreases when the band splits as a result of the application of $E_{\rm ext}$. In Fig.~\ref{fig:band_Efield}, we can see that states that are closer in energy at a given  {\bf k} point tend to display increased shared weight on the two layers of the bilayer, as is the case for the second and third valence states at $K$, $K'$, and $\Gamma$.

\begin{figure*}[t]
\centering
\includegraphics[width=0.65\textwidth] {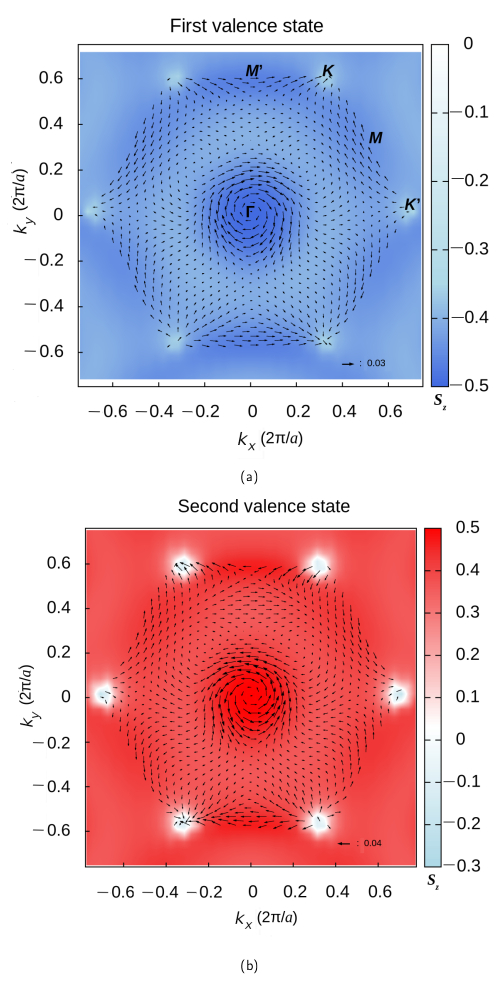}
\caption{\label{fig:ST} Spin-texture plots for (a) the highest and (b) the second-highest valence states in the Brillouin zone of the antiferromagnetic CrI$_{3}$ bilayer in the presence of an electric field of 0.05~V/\AA\  applied  perpendicularly to the bilayer. Each of the states is layer-locked on one layer of the bilayer (see also Fig.~\ref{fig:band_Efield}), the first state on $l_2$ and the second on $l_1$.  The in-plane spin components are represented by vectors and the perpendicular spin component is represented by the color scheme at the right of the plots.
  The reference value of the magnitude of the in-plane spin component  is shown at the bottom of the plots. } % caption for whole figure
  % \end{center}
%\end{center}
\end{figure*}

In Fig.~\ref{fig:ST}, we display the calculated in-plane spin texture for the two highest valence states, split upon the action of $E_{\mathrm{ext}}$. We also show in Fig.~\ref{fig:ST} the isovalue color map of the perpendicular spin component $S_z$.
As discussed above and based on  Fig.~\ref{fig:band_Efield},  the spin textures in Fig.~\ref{fig:ST} also correspond to the layer-resolved spin textures (save for the states at $K$, $K'$, and $\Gamma$); the first split state represents the spin texture  on $l_2$ and the second split state represents the spin texture  on $l_1$.  For the first state $S_{z}$ is always negative, and for the other state it is mostly positive.
It can be seen immediately that the helicities of the in-plane spin texture on the two layers are opposite, as we expected on the basis of Fig.~\ref{fig:magn_ky_kx}, and each of the layers is fully covered by only one of the two highest states.  This confirms our deduction from Fig.~\ref{fig:magn_ky_kx} that the AFM CrI$_{3}$ bilayer is indeed characterized by a Rashba-type in-plane spin texture in which the two highest valence  states with opposite in-plane helicity are layer-locked with nearly full spatial segregation of the wavefunctions on the two layers. The in-plane spins have the largest magnitude at  a radius of approximately $0.15\times 2\pi/a$ around the $\Gamma$ point. They also have a relatively large amplitude around the $M$ point along the $K$-$M$-$K'$ line and around the $M'$ point along the $M'$-$K'$ line.
   On the basis of the amplitude of the in-plane spin around $\Gamma$ in  Fig.~\ref{fig:ST}, the Rashba coefficient for the upper valence band of the AFM bilayer is estimated to be 0.6~eV\,\AA\ (see the supplementary material).

It can be seen in Fig.~\ref{fig:ST} that in the parts of the BZ where the in-plane components are the strongest (in particular the ring region around the $\Gamma$ point with radius of approximately $0.15 \times 2\pi/a$, and the regions around the $M$ and $M'$ points)  the $S_{z}$ component also has rather large values. The largest values of $S_z$ are at $\Gamma$  and at $M'$ along  the line $M'$-$K$. At the same time, at the $K$ and $K'$ points, the  $S_z$ component virtually vanishes and even reverses sign in Fig.~\ref{fig:ST}(b)---these are the same points where the states interact and lose the layer locking present everywhere else. We assign the appearance of $S_z$ with reverse sign at the $K$ and $K'$ points in the second state to the presence of a very small amount of Cr $d$ and I $p$ states with opposite spin, observable already in the monolayer degenerate bands at $K$ (see the Appendix).
Thus, upon formation of the AFM bilayer, the monolayer states,
each having both  positive and negative $S_z$ at $K$ and $K'$, will interact between the layers and result in the  mixed $S_z$ character at each layer observed in Fig.~\ref{fig:ST} at $K$ and $K'$.  More details are given in the Appendix, also for the $\Gamma$ point.

$E_{\mathrm{ext}} = 0.05$~V/\AA\ is small enough to induce a linear splitting of the two states in the highest valence band and a first-order effect in $E_{\mathrm{ext}}$ on the spin texture.
In the limit of vanishing $E_\mathrm{ext}$, the in-plane spin textures of the two states have exactly the same magnitudes and opposite chiralities. The application of the electric field along the positive $z$ direction slightly increases  the amplitude of the in-plane spin component for the second valence state, while there is a small reduction of the in-plane spin component for the first valence state.
For example, with $E_\mathrm{ext}=0.05$~V/\AA, the largest amplitude of the in-plane spin increases (decreases) by $0.004$ for the second (first) state with respect to its average of 0.038 for the two states.
The difference in the strength of the in-plane spin components increases as $E_\mathrm{ext}$ increases.
It could be envisioned that the electric field could be used to tune the relative strength of the in-plane spin texture of the two layers/states for the bands well separated in energy.  The in-plane spin textures with $E_\mathrm{ext}=0.1$~V/\AA\ and the corresponding band structure are reported   in the supplementary material. For zero and small $E_\mathrm{ext}$, the contribution to the spin texture of the two layers  cancels in the band, while in the presence of a large external electric field, there is a nonvanishing overall spin texture.

We have shown that there is a pronounced layer locking of the in-plane-canted spin for the two highest valence states in AFM bilayer CrI$_3$.
This layer locking without the typical Rashba energy splitting of the degenerate states has
some resemblance to the mechanism of the ``hidden Rashba spin-orbit splitting'' (R-2) in centrosymmetric nonmagnetic crystals \cite{YuaLiuZha19,ZhaLiuLuo14}, although in our case there is no (hidden) Rashba energy splitting between the two doubly degenerate bands interacting via the Rashba term  (see the supplementary material). In nonmagnetic centrosymmetric crystals, all bands must be at least doubly degenerate, and global inversion symmetry (I) does not allow Rashba splitting into singly degenerate bands of opposite spin polarization.
 However, the presence of local asymmetry in the individual spatial I-partner sectors does permit the R-2 effect (with splitting into doubly degenerate bands) when the doubly degenerate states with opposite spin on the different sectors are  prevented from mixing by a mechanism such as symmetry-enforced wavefunction segregation in nonsymmorphic crystals \cite{YuaLiuZha19}.
In the AFM bilayer, neither time-reversal symmetry nor inversion symmetry is present. All bands must, however, be doubly degenerate because of the presence of the symmetry operation  I$\cdot$T, and (although separated in energy) such bands interact via the Rashba term, as local asymmetry is also present in the individual I$\cdot$T-partner sectors. The resulting two degenerate wavefunctions of the highest valence band are almost completely spatially segregated  on the two layers (except near the BZ edge around the $K$ and $K'$ points, and at $\Gamma$)---see Fig.~S3 in the supplementary material. In the R-2 effect, the segregation is not always complete, and when it is complete (in nonsymmorphic crystals), it occurs only along the high-symmetry direction in the BZ, where the  symmetry enforcing the wavefunction segregation is present \cite{YuaLiuZha19}.
  In the AFM bilayer, an effect similar to (but not prohibitive for the mixing) that of the nonsymmorphic symmetries is produced by the layered  AFM order with a perpendicular easy axis that naturally segregates the wavefunctions on the two layers. However, in our case, at the {\bf k} points where already in the FM monolayer two states are degenerate or quasi-degenerate, a complete spatial segregation in the AFM bilayer does not occur (see Figs.~S5 and S6 in the supplementary material). Thus, we conclude  that for a successful spatial  wavefunction segregation of the two degenerate states of a van der Waals AFM bilayer band, it is crucial that the relevant  monolayer states as well as the AFM band in question remain well separated in energy from the other bands.

\section{Conclusions}
We have studied the spin texture in momentum space of the highest valence band of the AFM CrI$_{3}$ bilayer characterized by the perpendicular easy axis. We found that it displays a Rashba spin texture with opposite sign of the in-plane spins on the two layers. In this case, the Rashba effect is not accompanied by an energy splitting of the degenerate bands. We applied a small electric field perpendicular to the bilayer that splits the degenerate states according to their layer localization. We observed a layer segregation of the states over nearly the whole BZ, each state belonging to only one of the two layers. At the high-symmetry points $K$ and $K'$, and to a lesser extent at $\Gamma$, the segregation is destroyed by band interactions originating from degeneracies or near-degeneracies present in the CrI$_3$-monolayer band structure at those points. We compared the Rashba effect and layer locking of the in-plane-canted spin in the AFM CrI$_3$ bilayer with the R-2 effect and the hidden spin polarization in nonmagnetic centrosymmetric crystals. Although in both cases the bands must remain doubly degenerate and the Rasbha effect is enabled by the local asymmetry in the individual spatial inversion-related sectors, the segregation  mechanisms differ  in the two cases.  We conclude that, while the layered AFM coupling makes the spatial segregation possible, the necessary condition is the separation in energy of the band in question.

We expect that a similar effect should be observed in other van der Waals AFM bilayers featuring a perpendicular easy axis, strong spin-orbit coupling, and a band isolated in energy. In all such systems,  the relative  strength of the in-plane canting of the spins of the two states/layers could also be  tuned by the application of an external electric field.

\FloatBarrier

\setcounter{section}{0}

\renewcommand{\thesection}{A}
\renewcommand{\thesubsection}{A\arabic{subsection}}

\section{Appendix}

\setcounter{figure}{0}

%\subsection{\label{app:A1}Band structure of monolayer CrI$_{3}$}

%\renewcommand{\thesubsection}{A\arabic{subsection}}
%\renewcommand{\thefigure}{A\arabic{figure}}

\begin{figure*} [pos=t]
\centering
\includegraphics[width=0.75\textwidth] {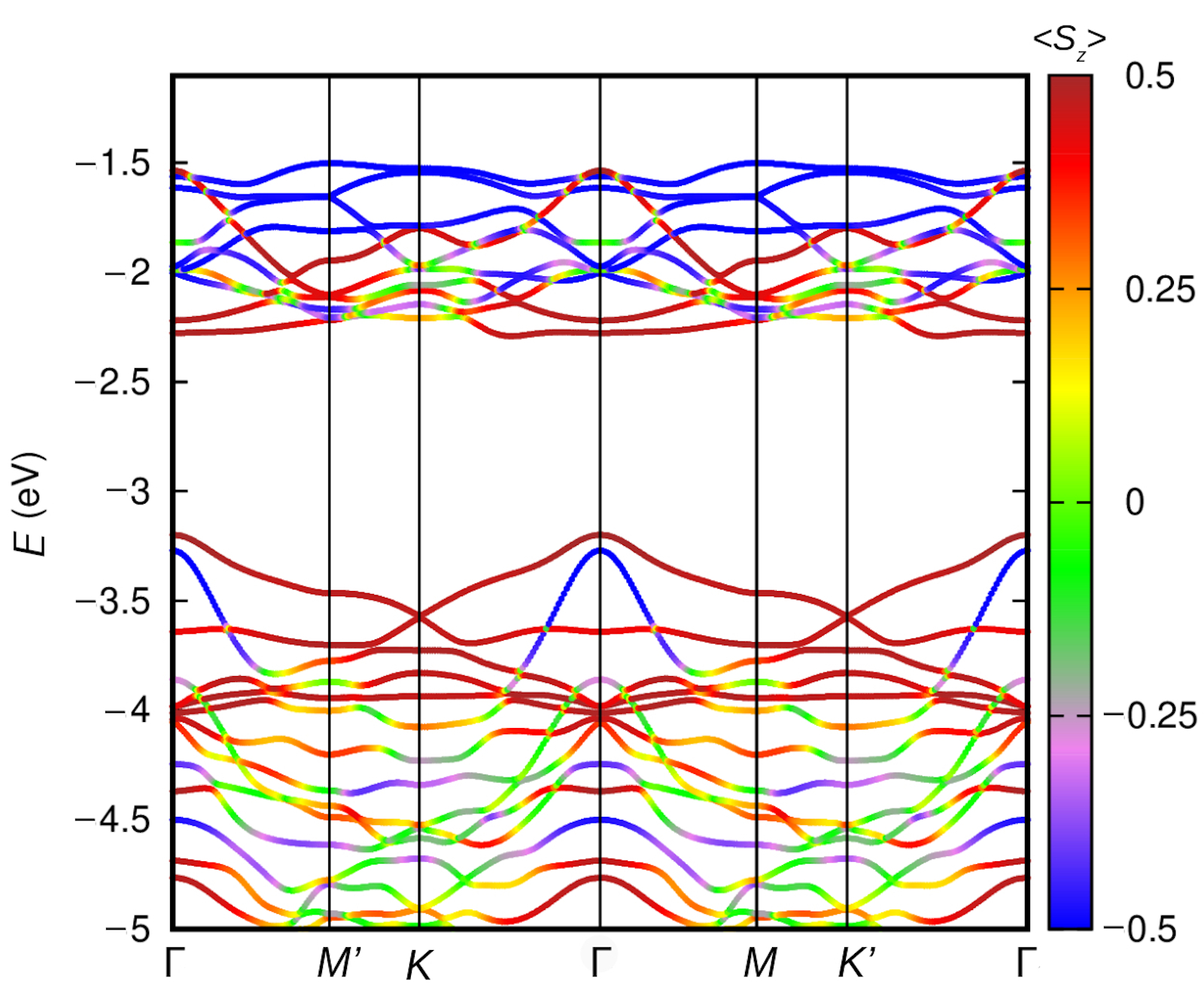}
\renewcommand{\thefigure}{A\arabic{figure}}
\caption{\label{fig:ML_Sz} Spin-projected band structure of the ferromagnetic CrI$_{3}$ monolayer. The projection of the  mean $S_{z}$ of the states is color-coded. The energy window
includes the highest valence bands and lowest conduction bands.}
\end{figure*}

In Figs.~\ref{fig:band_Efield} and \ref{fig:ST} we see that the layer locking of the two highest valence states shows some anomalies at the $K$ and $K'$ points, and to some extent also at $\Gamma$. This can be understood on the basis of the band structure of the pristine CrI$_{3}$ monolayer in the FM configuration, which is given in Fig.~\ref{fig:ML_Sz}. There, we also show the $S_{z}$ projection on the energy bands. We see that  the highest and second-highest valence states are degenerate at $K$  and that they are very close in energy at $\Gamma$. $S_{z}$ has the same sign for the first and second valence states at $K$, while at $\Gamma$ the two states have opposite signs.  The  whole highest valence band is of roughly the same $S_z$ (although it is somewhat reduced near $K$).

The band structure of the weakly interacting bilayer with the AFM arrangement of layers can be viewed as a superposition of the $S_{z}$-projected band structure (in  Fig.~\ref{fig:band_Efield}) and its time reversal ($-S_{z}$-projected band structure) with  some further modulation of the band-structure dispersion due to the weak interlayer interactions. Indeed, the AFM CrI$_3$ bilayer \cite{GhoStoBin21} has  dispersions in the BZ for the highest valence states rather similar to those of the FM monolayer from Fig.~\ref{fig:ML_Sz}.
The main difference is the increase in energy of the first band around the $\Gamma$ point and  removal of the band crossing at $K$ and $K'$ in the bilayer case due to the interaction between degenerate states of the two layers.

It is thus  expected that the character of the FM monolayer states at a given {\bf k} point  will determine to a good measure the character of the bilayer states formed from those monolayers. The highest  valence state of the CrI$_{3}$ monolayer will form the first valence band of the AFM bilayer, with twofold degeneracy coming from the two layers with opposite spins. Interestingly, the band crossing at $K$ of the two highest valence bands in the monolayer is nontrivial, as the highest valence  band
has a Chern number of 2 and the second-highest valence band near $K$  is characterized by a Chern number of $-2$ \cite{BaiYuKim18}. The 3D probability density of the monolayer's two highest states at $K$ (see Fig.~S6 in the supplementary material) shows that these states have localized density on different Cr atoms.
The mean $S_z$ for the degenerate states at $K$ in the monolayer is only approximately $0.34$. This makes it possible for the states of the two layers with opposite magnetization to hybridize upon formation of the bilayer, creating an energy splitting in the band structure of the bilayer.

At the $\Gamma$ point, the two highest valence states of the monolayer are of opposite spin and rather close in energy. This allows some interactions between the two states on the different layers, upon formation of the AFM bilayer. In this case, however, because of the energy difference, each state of the monolayer only weakly hybridizes with the state of the other layer.
%The other state from the highest valence AFM bilayer band of the opposite spin will originate from the interaction between the first monolayer state with the dominant spin on $l_2$ and the second monolayer state on $l_1$.

\section*{Declaration of competing interest}
The authors declare that they have no known competing financial interests or personal relationships that could have appeared to influence the work reported in this article.

\section*{Data availability}
The data that support the findings of this study are available from
the corresponding author upon reasonable request.

% Uncomment and use as the case may be
%\begin{theorem}
%\end{theorem}

% Uncomment and use as the case may be
%\begin{lemma}
%\end{lemma}


%\section{}\label{}


\printcredits

%\bibliographystyle{CrI3_AFM}

%\bibliographystyle{unsrtnat}

%\bibliography{CrI3_AFM}

\providecommand{\noopsort}[1]{}\providecommand{\singleletter}[1]{#1}%



\end{document}